\documentclass[review]{elsarticle}
\usepackage{amssymb,latexsym,amsmath}
\usepackage{graphicx,times}
\usepackage{dcolumn}
\usepackage[margin=1.0in]{geometry}
\newtheorem{thm}{Theorem}

\newtheorem{Def}{Definition}[section]

\numberwithin{equation}{section}
\journal{Journal of \LaTeX\ Templates}

\begin{document}

\begin{frontmatter}

\title{ Series Solution of System of Fractional Order Ambartsumian Equations: Application in Astronomy}

\author{Jayvant Patade\fnref{myfootnote} }
\ead{dr.jayvantpatade@gmail.com}
\address{Department of Mathematics, Jaysingpur College, Jaysingpur, Kolhapur - 416101, India.}


\cortext[mycorrespondingauthor]{Corresponding author}
\begin{abstract}
The Ambartsumian equation is arising in astronomy and used in the theory of surface brightness in a milky way. In this paper, we introduce system of fractional order Ambartsumian equations and use the Picard iterative method to obtain the series solution of these equations. The solution is provided in the form of a power series which is convergent for all reals. We prove the convergence of this series.
\end{abstract}
\end{frontmatter}

\section{Introduction}
\par In \cite{Milky} Ambartsumian derived a delay differential equation describing the fluctuations of the surface brightness in a milky way. The equation is described as:
\begin{equation}
y'(t) = -y(t)+\frac{1}{q}y\left(\frac{t}{q}\right)\label{1}
\end{equation}
where $q>1$ and is constant for the given model.\\
Recently, Patade and Bhalekar \cite{Patade} solved the Ambartsumian equation by
using the Daftarday-Gejji and Jafari technique. Kumar et.al \cite{Kumar} used the homotopy analysis transform method (HATM) and Khaled et.al  \cite{Khaled} the homotopy perturbation method (HPM) for solving a fractional form of the Ambartsumian equation. Ebaid, et.al \cite{Ebaid} used the Adomian decomposition method (ADM) to solve the Ambartsumian equation. Adel et.al \cite{Adel} have successfully employed a numerical method to solve the Ambartsumian equation. In this paper, we generalize the Ambartsumian equation (\ref{1}) to the system of fractional order Ambartsumian equations using the Caputo derivative.

\section{Preliminaries}\label{Pre}
In this section, we give some basic definitions and results regarding fractional calculus.

\begin{Def}\cite{Kilbas}
	The Riemann-Liouville (RL) fractional integral
	is defined as
	\begin{equation}
	I^\mu f(t)=\frac{1}{\Gamma(\mu)} \int_{0}^t (t-\tau)^{\mu-1}f(\tau)d\tau,\quad \text{if}\quad m-1<\mu<m, \quad m\in \mathbb{N}.
	\end{equation}
\end{Def}

\begin{Def}\cite{Kilbas}
	The Caputo fractional derivative is defined as:
	\begin{eqnarray}
	D^\mu f(t)&=&\frac{d^m}{ dt^m} f(t),\quad \mu = m \nonumber\\
	&=& I^{m-\mu}\frac{d^m}{ dt^m} f(t),\quad {m-1} <\mu <m,\quad m\in \mathbb{N}.
	\end{eqnarray}
\end{Def}
Note that for $0\le m-1 < \alpha \le m$ and $\beta>-1$
\begin{eqnarray}
I^\alpha x^\beta &=&\frac{\Gamma{(\beta+1)}}{ \Gamma{(\beta+\alpha+1)}} x^{\beta+\alpha},\nonumber\\
\left(I^\alpha D^\alpha f\right)(t)&=& f(t)-\sum_{k=0}^{m-1} f^{(k)}(0)\frac{t^k}{k!}.
\end{eqnarray}

\begin{Def}\cite{Kilbas}
	The  Mittag-Leffler function is defined as  
	\begin{equation}
	E_\alpha (x)=\sum_{n=0}^\infty\frac{x^n}{\Gamma{(\alpha n+1)}}, \quad \alpha>0.
	\end{equation}
\end{Def}

\section{ System of Fractional Order Ambartsumian Equations}
In this section, we generalize the Ambartsumian equation (\ref{1}) to the system of fractional order Ambartsumian equations as: 

\begin{equation}
D^\alpha y(t) = -Iy(t)+By\left(\frac{t}{q}\right),  \quad y(0)=\lambda, \quad 0< \alpha \le 1, \label{2}
\end{equation}
where  $D^\alpha$ denotes Caputo fractional derivative, $I$ is the identity matrix of order $n$,  $1<q$, \\
$y=$$
\begin{bmatrix}
y_1\\
y_2\\
\vdots\\
y_n
\end{bmatrix}
$, 
$\lambda=$
$\begin{bmatrix}
\lambda_1\\
\lambda_2\\
\vdots\\
\lambda_n
\end{bmatrix}$
and 
 $B=$$
\begin{bmatrix}
\frac{1}{q}& 0 & 0 & \cdots & 0\\
0& \frac{1}{q} & 0 & \cdots & 0\\
\vdots& \vdots & \vdots & \ddots & 0\\
0& 0 & 0 & \cdots & \frac{1}{q}
\end{bmatrix}_{n\times n}.
$\\

 Applying Picard iterative method to the initial value problem (\ref{2}), we have
 \begin{equation}
  y(t) = y(0) -I J^\alpha y(t)+ B J^\alpha y\left(\frac{t}{q}\right)\label{3}.
 \end{equation}
 Suppose $\phi_k(t)$ be the $k^{th}$ approximate solution, where the initial approximate solution is
 taken as
\begin{equation}
\phi_0(t)=\lambda.
\end{equation}

For $k \ge 1$, the recurrent formula as below:
 \begin{equation}
\phi_k (t) =\lambda -I J^\alpha \phi_{k-1} (t)+ B J^\alpha  \phi_{k-1}\left(\frac{t}{q}\right)\label{4}.
\end{equation}

From the recurrent formula, we have

\begin{eqnarray}
	\phi_1 (t) &=& \lambda -I J^\alpha \phi_{0} (t)+ B J^\alpha  \phi_{0}\left(\frac{t}{q}\right)\nonumber\\
	&=&\lambda -I \frac{\lambda t^\alpha}{\Gamma (\alpha+1)}+ B \frac{\lambda t^\alpha}{\Gamma (\alpha+1)}\nonumber\\
	&=&\left(I +(-I +B)\frac{ t^\alpha}{\Gamma (\alpha+1)}\right)\lambda,\nonumber\\
\phi_2 (t) &=& \lambda -I J^\alpha \phi_{1} (t)+ B J^\alpha  \phi_{1}\left(\frac{t}{q}\right)\nonumber\\
&=&\lambda -I J^\alpha \left[\left(I +(-I +B)\frac{t^\alpha}{\Gamma (\alpha+1)}\right)\lambda\right]+ B J^\alpha  \left(I +(-I +B)\frac{ q^{-\alpha}t^\alpha}{\Gamma (\alpha+1)}\right)\lambda\nonumber\\
&=&\lambda -I \left[\frac{\lambda t^\alpha}{\Gamma (\alpha+1)} +(-I +B)\frac{\lambda t^{2\alpha}}{\Gamma (2\alpha+1)}\right]+ B   \left[\frac{\lambda t^\alpha}{\Gamma (\alpha+1)} +(-I +B)\frac{\lambda q^{-\alpha} t^{2\alpha}}{\Gamma (2\alpha+1)}\right]\nonumber\\
&=&\left[I +(-I +B)\frac{ t^\alpha}{\Gamma (\alpha+1)} +  (-I +Bq^{-\alpha})(-I +B)\frac{ t^{2\alpha}}{\Gamma (2\alpha+1)}\right]\lambda,\nonumber\\
\phi_3 (t)&=&\left[I +(-I +B)\frac{ t^\alpha}{\Gamma (\alpha+1)} +  (-I +Bq^{-\alpha})(-I +B)\frac{ t^{2\alpha}}{\Gamma (2\alpha+1)}\right.\nonumber\\
&&\left.+(-I +Bq^{-2\alpha})(-I +Bq^{-\alpha})(-I +B)\frac{ t^{3\alpha}}{\Gamma (3\alpha+1)}\right]\lambda,\nonumber\\
&&\cdots,\nonumber\\
\phi_k (t)&=& \left[I + \sum_{m=1}^k\prod_{j=1}^{m}(-I+Bq^{-(m-j)\alpha})\frac{ t^{m\alpha}}{\Gamma (m\alpha+1)}\right]\lambda\nonumber
\end{eqnarray}
As $k\rightarrow\infty$,$\quad \phi_k (t)\rightarrow y(t)$
\begin{eqnarray}
y(t)&=& \left[I + \sum_{k=1}^\infty\prod_{j=1}^{k}(-I+Bq^{-(k-j)\alpha})\frac{ t^{k\alpha}}{\Gamma (k\alpha+1)}\right]\lambda.\nonumber
\end{eqnarray}
If we set $\prod_{j=1}^{k}(-I+Bq^{(k-j)\alpha})=I$, for $k=0$,
then

\begin{eqnarray}
 y(t)&=& \left[\sum_{k=0}^\infty\prod_{j=1}^{k}(-I+Bq^{-(k-j)\alpha})\frac{ t^{k\alpha}}{\Gamma (k\alpha+1)}\right]\lambda.\label{5}
\end{eqnarray}

\begin{thm}
	 For $q>1$, the  power series 
	\begin{eqnarray}
	y(t)&=& \left[\sum_{k=0}^\infty\prod_{j=1}^{k}(-I+Bq^{-(k-j)\alpha})\frac{ t^{k\alpha}}{\Gamma (k\alpha+1)}\right]\lambda\nonumber
	\end{eqnarray}
 is  convergent for $t\in\mathbb{R}$.
\end{thm}
\textit{\textbf{Proof:}}$\quad$
Result follows immediately by ratio test \cite{Apostal}.

\textbf{Inparticular:}\begin{eqnarray}
\begin{bmatrix}
D^\alpha y_1(t)\\
D^\alpha y_2(t)
\end{bmatrix} &=& \begin{bmatrix}
-y_1(t) & 0\\
0 & -y_2(t)
\end{bmatrix} +\begin{bmatrix}
\frac{1}{q}y_1\left(\frac{t}{q}\right) & 0\\
0 & \frac{1}{q}y_2\left(\frac{t}{q}\right)
\end{bmatrix} , \nonumber\\
&&  \begin{bmatrix}
y_1\\
y_2
\end{bmatrix}=\begin{bmatrix}
\lambda_1\\
\lambda_2
\end{bmatrix}, \quad 0< \alpha \le 1, \quad q>1.\label{6}
\end{eqnarray}
The solution of Eq. (\ref{6}) is 
\begin{eqnarray}
y_1(t)&=& \left[\sum_{k=0}^\infty\prod_{j=1}^{k}(q^{-(k-j)\alpha-1}-1)\frac{ t^{k\alpha}}{\Gamma (k\alpha+1)}\right]\lambda_1,\label{7}\\
y_2(t)&=& \left[\sum_{k=0}^\infty\prod_{j=1}^{k}(q^{-(k-j)\alpha-1}-1)\frac{ t^{k\alpha}}{\Gamma (k\alpha+1)}\right]\lambda_2.\label{8}
\end{eqnarray}

\section{Conclusions}\label{concl}
 The fractional delay differential equations are helpful in model memory phenomena and hereditary properties. In general, it is not possible to find exact solutions of these equations. In this case, the methods like Picard iterative method are proved useful in producing at least approximate analytical results. In this work, we were able to find an exact solution of the system of fractional order Ambartsumian equations by using the Picard iterative method. It is shown that the solution series is convergent for all values of reals and hence it can be used as a global solution.

\subsection{Acknowledgements}

The author acknowledges the Savitribai Phule Pune University, Pune, India for the postdoctoral fellowship [SPPU-PDF/ST/MA/2019/0001].

\end{document}